\documentclass{article}
\usepackage{emulateapj}
\def\solar{\ifmmode_{\mathord\odot}\;\else$_{\mathord\odot}\;$\fi}
\begin{document}

\slugcomment{To appear in The Astrophysical Journal}
\title{\center Infalling Gas Towards the Galactic Center}
\author{Alison L. Coil and Paul T.P. Ho}
\affil{Harvard-Smithsonian Center for Astrophysics \\
60 Garden St. MS-78\\
Cambridge, MA 02138\\
e-mail: acoil@cfa.harvard.edu; pho@cfa.harvard.edu \\}

\begin{abstract}

\ \ \ \ \ VLA maps of ammonia emission were made for the 
Galactic Center region.  The NH$_3$(1,1) and NH$_3$(2,2) transitions 
were observed 
in three 2$'$$\times$2$'$ fields covering Sgr A$^*$ 
and the region 3$'$ immediately south of it.  In the central 3 
parsecs surrounding 
Sgr A$^*$ we find emission which appears to be associated with the 
circumnuclear 
disk (CND), both morphologically and kinematically.  This central emission
is connected to a long, narrow 2 pc$\times$10 pc streamer of clumpy
molecular gas located towards the south, which appears to be carrying 
gas from the nearby 
20 km sec$^{-1}$ giant molecular cloud (GMC) to the circumnuclear region.  
We find a velocity gradient along the streamer, with progressively higher
velocities as the gas approaches Sgr A$^*$. 
The streamer stops at the location of the CND, where the line 
width of the NH$_3$ emission increases dramatically.
This may be the kinematic signature of accretion onto the 
CND.  The ratio of the NH$_3$(2,2)/NH$_3$(1,1) emission indicates that
the gas is heated at the northern tip of the streamer, located inside the
eastern edge of the CND.  The morphology, kinematics and temperature 
gradients of the gas all indicate that the southern streamer is located at
 the Galactic Center and is interacting with the circumnuclear region.
\end{abstract}


\section{Introduction} 

\ \ \ \ \ Our Galactic Center has been under increasingly intensive study 
of late as more sensitive instruments with higher angular resolutions are 
brought to bear on the region.  Recent data 
suggest that many galactic nuclei harbor massive black holes (\cite{kor95}; 
\cite{fer96}; \cite{kor96}; \cite{kom96}; \cite{van97}), and there is 
mounting evidence that Sgr A$^*$, the strong
compact radio source at the dynamical center of the Milky Way, is also a
black hole with a mass of $\sim$2 10$^6$ M$\solar$ (\cite{mor96};
\cite{eck97}; \cite{gen97}). 
Immediately surrounding Sgr A$^*$, arc-shaped ionized gas streamers 
converge at the location of the proposed black hole.  These streamers
may originate from the inner edge of the circumnuclear disk (CND) 
 (\cite{gen85}; \cite{har85}; \cite{ser86}; \cite{gat86}; \cite{gus87}; 
\cite{sut90}; \cite{mar93}; \cite{mar95}), where the gas
on the inner edge is photoionized and stripped off the disk, falling in 
towards the nucleus.  The CND is dense (10$^5$ cm$^{-3}$), clumpy, 
and turbulent with large line widths ($\geq$40 km sec$^{-1}$).  
Its inner edge is located 1.5 - 2 pc from Sgr A$^*$, and it extends 
beyond 7 pc.  The gas is predominantly moving in a circular orbit,
rotating around the nucleus at a velocity of $\sim$110 km sec$^{-1}$.  
However, non-circular motions are also present, and as the disk appears to be
only semi-complete and the gas is highly clumped, the CND does not appear to be 
in equilibrium.  The CND may be feeding the nucleus through 
the ionized streamers (\cite{gus87}), which might be infall from the CND.
The question remains as to how the disk itself is fed. \newline

\ \ \ \ \ Two giant molecular clouds (GMCs) lie near the Galactic Center, 
the ``20 km sec$^{-1}$ cloud'' (M-0.13-0.08) and the ``50 km sec$^{-1}$ cloud'' 
(M-0.02-0.07) (\cite{gus81}).  These clouds are physically near the Galactic Center, 
and various studies have attempted to establish the 
interactions of the GMCs, the continuum Sgr A gas complex, and the 
CND and ionized gas streamers surrounding 
Sgr A$^*$ (\cite{hoe85}; \cite{lis85}; \cite{oku89}; \cite{gen90}; Zylka, 
Mezger, \& Wink 1990; \cite{hoe91}; \cite{ser92}; \cite{den93}; \cite{mar95}).
  Ho et al. (1991) map in NH$_3$(3,3) a long gas streamer 
which connects the 20 km sec$^{-1}$ cloud to the location of the 
CND.  They propose that this gas streamer feeds the circumnuclear region 
with gas from the 20 km sec$^{-1}$ cloud.  No velocity 
gradient along the streamer is detected within their spectral
resolution of 10 km sec$^{-1}$, and no increase in line width is seen, 
prompting an interpretation that the gas is moving towards Sgr A$^*$ across 
the line of sight. Hints of this streamer are also seen in HCN 
(Marr et al. 1995) and submillimeter continuum emission (\cite{den93}).
However, these studies also find no clear velocity gradient in the streamer 
along the line of sight. \newline

\ \ \ \ \ In this paper we investigate the interactions between the 20 km 
sec$^{-1}$ cloud and the circumnuclear region in order to question how gas is 
drawn into the CND.  We also compare the HCN information on the CND with 
our NH$_3$ findings.  Section 2 of this paper 
details our NH$_3$(1,1) and NH$_3$(2,2) observations and data reduction.  
The results are presented in Section 3 and discussed in Section 4, while
 Section 5 outlines our conclusions.  \newline

\section{Observations and Data Reduction}

\ \ \ \ \ We observed the (J,K)=(1,1) and (2,2) metastable transitions of 
NH$_3$ emission on the night of 
February 10 of 1995 with the Very Large Array (VLA) in
C/D configuration.\footnote{The VLA is operated by NRAO and the Associated 
Universities Inc., under cooperative agreement
with the National Science Foundation.}  Our observing frequencies were
 23.694495GHz and 23.722633GHz.  We made three pointings with one field 
centered at ($\alpha$, $\delta$)=(17$^h$42$^m$29$^s$.3, 
-28$^\circ$$59'$$17''$.0), the position of Sgr A$^\ast$, and two fields
centered at 
($\alpha$, $\delta$)=(17$^h$42$^m$29$^s$.3, -29$^\circ$$00'$$18''$.6)
and ($\alpha$, $\delta$)=(17$^h$42$^m$29$^s$.3, -29$^\circ$$02'$$18''$.6), 
roughly one and three arcminutes south of the first field.  Our
total spectral bandwidth is 12 MHz, centered at $v_{LSR}$= -10.56 km 
$s^{-1}$ for the northern and central field and $v_{LSR}$= 31.14 km 
$s^{-1}$ for the southern field.  At
$\lambda$=1.3 cm the field-of-view is $2'$ as determined by the
primary beam of the individual antennas. The full
resolution of our synthesized images is $\sim$$3''.6$$\times$$2''.1$
with a position angle of 44$^{\circ}$.  Our spectral resolution is 
4.9 km sec$^{-1}$.  \newline

\ \ \ \ \ We reduced the data within AIPS, flagging potentially bad data 
surrounding the amplitude and phase discontinuities in
the calibrators.  Our flux calibrator was 3C286, and we tracked  
the antenna gain and phase responses with the phase
calibrator 1730-130.  After calibrating the 
broadband data, we applied 
the calibration tables to the line data, using the strong quasar 
1226+023 as our bandpass calibrator.  The continuum emission in 
the northern two fields was subtracted from the line data by averaging
the {\it u,v} data in several off-line channels and
subtracting this from the {\it u,v} data of all of the line channels.  
\newline

\ \ \ \ \ The data were imaged and analyzed with the MIRIAD software package. 
To emphasize the extended emission in our images we heavily tapered 
the {\it u,v} data by convolving with a Gaussian 
function, resulting in
 a beam size of $\sim$14$''$$\times$9$''$ with a position angle of 
12$^{\circ}$. 
In order to avoid CLEANing regions of the ``negative bowl'' artifically 
caused by the lack of short spacing information in the {\it u,v} plane 
and to help the deconvolution algorithm interpolate through the central
hole in the {\it u,v} plane, we added a flat offset from zero of 50 mJy
 across all of the spectral images to bias the CLEANing towards positive
emission.  This flat offset was subtracted out after CLEANing, and we compared 
our data cube with images made without zero spacing flux to 
check that source structure did not change as a result of this process 
and did not depend on the amount of zero
spacing flux introduced.  A variation of this procedure using AIPS is
described in detail by 
Wiseman (1991) and Wiseman \& Ho (1998) who provide sample
maps made with different amounts of added zero spacing flux as well as with 
different deconvolution techniques.  This process
does not intend to correct for the true value of zero spacing flux in
the source, and as a result we estimate that as much as 1/2 to 2/3 of the 
total flux may be missing in our maps due to extended structure not adequately
 sampled because of the lack of short spacing data. \newline

\ \ \ \ \ The {\it u,v} data from the northern two fields were imaged and 
deconvolved simultaneously as one image with two pointings.  When performing
a joint mosaic, the imaging software in MIRIAD corrects for primary beam
attenuation to within a certain noise limit, as set by the theoretical noise 
of the individual
pointings.  As a result, while most of the image is primary beam corrected,
there is residual attenuation at the edges of the mosaic.  The 
far southern field has a different v$_{LSR}$ and an offset velocity
coverage from the northern two fields, so the data from the southern field 
were imaged and deconvolved separately.  To mosaic the subset of the data with 
overlapping velocity channels in all three fields, we de-mosaiced the northern
two fields using the task DEMOS and then applied a linear mosaic of all three
fields in the image plane using the task LINMOS with the taper option,
which creates final images which have significant attenuation at the edges 
(\cite{sau96}).  \newline

\section{Results}

\subsection{Velocity-Integrated Map}

\ \ \ \ \ Figure \ref{fig.color.mom0} presents the velocity-integrated 
NH$_3$(1,1) and NH$_3$(2,2) emission from all three fields in contours 
overlaid on a false color HCN map of the inner part of the CND.  The
scale of this image is $\sim$9 pc$\times$16 pc (using R$_\circ$=8.5 kpc),
and there is significant attenuation at the edges of this mosaic so that only
the highest signal-to-noise features are seen here. 
The HCN emission (\cite{mar93}) 
has a beam size of 12$''$$\times$6$''$, while the NH$_3$ line emission has a 
beam size of 14$''$$\times$9$''$, as seen in the lower left corner of each 
image.  A long narrow north-south streamer can be seen 
in both NH$_3$ maps.  The southern half of this streamer extends into the 
20 km sec$^{-1}$ GMC, and the streamer ends to the north at the location 
of the CND, precisely where little or no HCN emission is 
seen (\cite{mar93}; \cite{jac93}). 
A semi-complete ring is mapped in NH$_3$(1,1) around the 
CND, extending roughly 2-4 pc from Sgr A$^*$.  The NH$_3$(2,2)
map shows a less-complete ring of emission which closely follows the 
curve of the HCN emission on the eastern and western sides. 
Most of the northern half of the CND is missing in the NH$_3$ maps, and
this is almost certainly due to the limited spectral window of this experiment.
The HCN data show that emission in the northeastern quadrant is predominantly
at v$_{LSR}$$\sim$100 km sec$^{-1}$, which falls outside of our velocity coverage. 
\newline

\ \ \ \ \  The size of the entire southern streamer connecting
 the 20 km sec$^{-1}$ GMC to the circumnuclear region is about 
$\sim$2 pc$\times$10 pc, with many smaller clumps about 1 pc to 2 pc in
diameter.  It is immediately clear from Figure \ref{fig.color.mom0}
that the greatest morphological differences between the NH$_3$(1,1) and 
NH$_3$(2,2) emission is in the immediate neighborhood of Sgr A$^*$ and the
CND. The NH$_3$(1,1) emission is fluffier than the NH$_3$(2,2), extending further out
from the CND as defined by the HCN emission.  The NH$_3$(2,2) gas
appears in projection to be concentrated near the inner edge of the disk.  
In both maps the streamer 
has a large clump in the north where the streamer meets the disk.
The northern tip of the streamer extends further into the interior of the 
disk in the NH$_3$(2,2) map.  Both maps also show a dense region of gas
along the western side of the disk, near the densest part of the CND itself.
The NH$_3$(1,1) emission continues on around the ring to the northwest 
at a larger radial distance before sharply stopping where the CND has 
two dense spots, one to the north of the other, where little NH$_3$(2,2) emission 
is seen.  \newline

\ \ \ \ \ Figure \ref{fig.bw.mom0.spiral} overlays the same NH$_3$(1,1) 
and NH$_3$(2,2)
contours from the northern half of the streamer onto a greyscale map of the 
ionized gas streamers which form a mini-spiral centered at Sgr A$^*$ 
(\cite{loe83}).
It has been proposed that the western streamer is the inner photo-ionized
edge of the CND and that the northern and eastern streamers are
infalling gas which has been stripped off of the disk (\cite{lac91}).  The 
northern part of the streamer reaches up past the eastern arm of the mini-spiral.
In the NH$_3$(2,2) map the small clump at the tip of the streamer
resides in the hollow area between the eastern and northern ionized
gas streamers. This is the same location as that of the ionized gas and dust
`tongue' which is thought to
be falling in towards Sgr A$^*$ (\cite{cha97} and references therein). 
On the western side of the spiral the NH$_3$(2,2) 
emission follows the outer edge of the western arm, which is also
the location of the inner edge of the CND.  In the NH$_3$(1,1)
map the northern part of the emission around the CND reaches down
in between the western and northern ionized gas streamers.
We find little emission in NH$_3$(1,1) and NH$_3$(2,2) in the central 
parsec of the Galaxy.  \newline

\subsection{Velocity-Integrated Map of the NH$_3$(2,2)/NH$_3$(1,1) Ratio} 

\ \ \ \ \ The rotation temperature of the gas can be determined 
by comparing the observed brightness temperatures of the two 
NH$_3$ transitions, using the optical depth measured in the NH$_3$(1,1) line.
For a given optical depth, a higher ratio of the NH$_3$(2,2) to
NH$_3$(1,1) observed brightness temperatures implies hotter gas.  
While in section 3.6 we calculate the rotation temperature for specific 
locations along the streamer where we can directly measure the optical depth 
of the NH$_3$(1,1) gas, here we present a spatial
representation of the NH$_3$(2,2) to NH$_3$(1,1) line ratio seen in this region. 
In Figure \ref{fig.22.11mom0} the contours map the NH$_3$(2,2) 
velocity-integrated emission in the northern part of the streamer.  This image
has been primary-beam corrected to within the theoretical noise limits, and
we present this map in order to investigate the
low-level emission around the CND, where the signal to noise is not as high as in
Figure \ref{fig.color.mom0}.   The NH$_3$(2,2)/NH$_3$(1,1) line ratio is shown
in greyscale emission, with the dark regions tracing higher ratios where the gas is
hotter. One can see that
the northern tip of the streamer is the darkest area on the map,
indicating that this gas is heated near the nuclear 
region.  There are also dark regions around the inner edge of the
northwestern part of the CND, and a somewhat dark area 
in between the two largest clumps in the streamer.  Inter-clump heating of gas
is commonly seen, where the gas is externally heated (\cite{wis98}). 
Figure \ref{fig.22.11heat} presents the same greyscale map of the 
NH$_3$(2,2)/NH$_3$(1,1) line ratio overlaid with contours of both the 
ionized gas
mini-spiral and the disk as mapped in HCN.  The heated tip of the NH$_3$
streamer lies in between the eastern and northern ionized 
gas streamers in the mini-spiral and is clearly located in projection 
inside the CND, near the inner northeast edge of the disk.  There is
another dark area of heated emission near the northeastern side of the CND,
though the signal-to-noise is lower in this area of the map. \newline

\subsection{Spectra}

\ \ \ \ \ To investigate the kinematics of the gas surrounding the CND, 
Figure \ref{fig.11.spec} presents selected NH$_3$(1,1) spectra.
The NH$_3$(1,1) velocity-integrated map shown here only includes
the northern two fields and has been corrected for primary beam 
attenuation so that the signal-to-noise is lower at the edges. We use
this primary-beam corrected map in order to probe the fainter emission 
around the nuclear region, and we show spectra of several interesting 
features around the CND.  
The semi-complete ring of NH$_3$(1,1) emission is clearly seen in 
the upper part of the map, while the southern half of the streamer is cut off to
the south due to the edge of the primary beam in the lower field.  We note that
in removing the continuum emission, we subtracted the {\it u,v} data
averaged over the velocity range of -35 km sec$^{-1}$ 
to -75 km sec$^{-1}$ from all the spectral channels.  It is possible 
therefore that our resulting
line-only data may be missing some emission in this velocity range. \newline

\ \ \ \ \ One can see from the spectra that the highest flux and 
signal-to-noise are found along the southern streamer.  Spectra J, K 
and L reveal that the gas in the streamer is centered around a
velocity of 25 km sec$^{-1}$, with FWHMs of 30-40 km sec$^{-1}$.  These 
three FWHM measurements may be overestimated due to
blending with hyperfine structures, which we will discuss shortly when we 
present the position-velocity diagrams in Section 3.4. 
In this figure, as we approach the CND in spectrum M, the 
gas becomes much more spread out in velocity space, with 
a FWHM $\geq$50 km sec$^{-1}$.  All of the other spectra which are near the 
CND or within 1$'$ of Sgr A$^*$ show emission with very large FWHMs. 
It is quite likely that some of the linewidths in the nuclear region 
are underestimated in our experiment due to our limited velocity range. 
HCN spectra from the CND show that in the southwest
region lines peak from -60 km sec$^{-1}$ to -120 km 
sec$^{-1}$, while in the northeast lines peak from 80 km sec$^{-1}$ 
to 110 km sec$^{-1}$.  These velocities are predominantly outside of our 
observed range, and this appears to be the principal reason why our data 
do not completely trace the CND. However, HCN spectra from these regions
also include broad wings of emission at higher and lower velocities,
and our spectra show emission from these broad wings. \newline

\ \ \ \ \ NH$_3$(2,2) spectra from the two northern fields are presented in 
Figure \ref{fig.22.spec}.
Here again the strongest emission is in the southern streamer, where 
the central velocity feature is around 20-35 km sec$^{-1}$.  As also seen in the
 NH$_3$(1,1) emission, NH$_3$(2,2) spectra near the nuclear region show an excess of  
emission with broad linewidths.  Most of the FWHM measurements in the NH$_3$(2,2)  
spectra are not affected by hyperfine blending, as the satellite lines 
 are resolved with our spectral resolution.   A dramatic
increase in the FWHM of the spectra can easily be seen near the CND.  To 
investigate low-level emission we present spectra 
from features at one and two contour levels in the velocity-integrated images.  
Features seen at two contour levels seem to have more significance
than those at one contour, based upon their spectra.   \newline

\ \ \ \ \  In order to verify that our continuum subtraction has 
been successful and that the passband correction is reliable, we present
spectra in Figure \ref{fig.base.spec} taken from randomly-selected areas 
where there is no emission seen in the velocity-integrated maps.  These 
spectra, plotted to the same scale as in Figures \ref{fig.11.spec} and \ref{fig.22.spec}, 
do not show any residual emission, thus it appears that our continuum subtraction
was successful. \newline

\subsection{Position-Velocity Diagrams}

\ \ \ \ \ Position-velocity diagrams 
are presented in Figure \ref{fig.psvl}.  The locations of three 
position-velocity cuts are shown in the NH$_3$(2,2) velocity-integrated map
at the left. The diagrams for these cuts are presented for 
both the NH$_3$(1,1) and NH$_3$(2,2) data, along with the NH$_3$(2,2)/NH$_3$(1,1) 
line ratio to examine the kinematics of the heated gas.  
Position-velocity diagrams for cut `a' taken north to south along 
the northern half of the streamer are presented in the top two rows,
where the second row of diagrams has been smoothed in the R.A. direction
to bring out the extended, tenuous emission. The top `a' row and the diagrams
for cuts `b' and `c' have had no smoothing applied.  \newline

\ \ \ \ \ In both the smoothed and unsmoothed data for cut `a', one can 
see a dramatic increase in linewidth in the upper half of the diagrams.
The gas approaches the nuclear region at 30-35 km sec$^{-1}$ with narrow 
linewidths, $\sim$10-15 km sec$^{-1}$ as seen in the NH$_3$(2,2) data.  
As mentioned before, the NH$_3$(1,1) emission has a hyperfine
line 7 km sec$^{-1}$ away from the main line which causes artificial line
broadening. The NH$_3$(2,2) emission more accurately reflects the inherent 
linewidths of the gas, as the hyperfine structures are clearly resolved. 
Gas in the northern tip of the streamer at the location of the CND
has very large linewidths, with emission seen across the entire sampled 
velocity range from 60 km sec$^{-1}$ to -25 km sec$^{-1}$. This extended emission
is more clearly seen for the NH$_3$(2,2) transition than the NH$_3$(1,1).
The smoothed NH$_3$(2,2)/NH$_3$(1,1) position-velocity diagram indicates 
that the warmer part 
of the gas (dark in the greyscale image) with large line
widths is at negative velocities from 0 km sec$^{-1}$ to -20 km sec$^{-1}$. 
In the unsmoothed `a' 
diagrams the region in between the two dense clumps of gas in the
streamer is seen at a higher flux level in the NH$_3$(2,2) data.  
This is reflected in the NH$_3$(2,2)/NH$_3$(1,1) 
divided position-velocity diagram where there is a dark region, indicating 
heated gas, at that location. In the smoothed  
NH$_3$(2,2)/NH$_3$(1,1) diagram this region is not as dark, probably because 
cooler gas surrounding this narrow region of heating has been included in the 
smoothed diagram.  \newline

\ \ \ \ \ The next row shows diagrams from cut `b' taken northeast to southwest along 
the southern part of the streamer.  An overall redshift can be seen in both 
transitions, moving from the southern part of the streamer up to the north.  
 In the NH$_3$(2,2) 
diagram the southern part of the streamer connects with the northern section
 at a velocity of 30 km sec$^{-1}$.  This connection is stronger
in the NH$_3$(2,2) image than in the NH$_3$(1,1) diagram, as was seen in the 
velocity-integrated map Figure \ref{fig.color.mom0}, where this inter-clump gas 
may be externally heated. At the lower edge of 
the diagram a `C' structure is visible in the gas, where the central region
of the cloud is blue-shifted along the line of sight compared to the 
northern and southern parts of the cloud. This feature can also be seen
in the next row of diagrams. \newline

\ \ \ \ \ The bottom row is from cut `c' taken east to west across the southern edge of 
the streamer, in the 20 km sec$^{-1}$ GMC.  
The `C' feature is prominent here in both the NH$_3$(1,1) and 
NH$_3$(2,2) diagrams. This may be indicating expansion where the central portion
of the cloud is blue-shifted towards the observer relative to the gas
to the east and west of it.  The NH$_3$(2,2)/NH$_3$(1,1) diagram shows that the
blue-shifted gas in the center of the cloud and the gas at the eastern edge
of the cloud is hotter than the gas along the western edge.  
Ho et al. (1985) suggest that a
supernova remnant may lie in the region south of the Sgr A East continuum
emission (roughly at RA=17$^h$42$^m$30$^s$ and Dec=-29$^\circ$$03'$), just east
of the bottom of the streamer.
They find evidence for an interaction such as a shock front, which could be
propagating into the molecular gas.  Our position-velocity diagrams are
consistent with this proposed idea.  The gas in this 
region is clearly being disrupted and heated, as our position-velocity diagrams show,
 and the gas may become destabilized by the disruption. The 
immense gravitational pull of Sgr A$^*$ and the central stellar cluster
 would then be able to strip the gas off
of the 20 km sec$^{-1}$ cloud and pull it in towards the nucleus along the
streamer.  \newline

\ \ \ \ \ Hyperfine structures are clearly present in these position-velocity
diagrams. While the first hyperfine line for NH$_3$(1,1) emission is blended 
with the central line, the second hyperfine line is located 
19 km sec$^{-1}$ away from the central line and can be seen in these diagrams, 
most clearly in the `a'  and `c' cuts. Hyperfine lines for the NH$_3$(2,2) 
transition are located 17 km sec$^{-1}$ and 26 km sec$^{-1}$ away from the 
central line and can be seen in the `c' (and to some extent in the `b') diagrams. 
Hyperfine features are also apparent in the spectra presented earlier (Figures 
\ref{fig.11.spec} and \ref{fig.22.spec}).  \newline

\subsection{Velocity Dispersion Map}
\ \ \ \ \ Figure \ref{fig.mom2} presents velocity dispersion maps for 
both the NH$_3$(1,1) and NH$_3$(2,2) emission.  These maps indicate the dispersion
of the gas around the central velocity feature and therefore trace
the line width of the emission, where the darker regions have greater line
widths.  In both transitions the FWHM of the gas 
is largest in the northernmost area around the CND, where emission from both the
western and eastern sides of the CND has large line widths. This same increase 
in line width in the nuclear region was seen in both the spectra and 
position-velocity diagrams.  Here we present a spatial map of this 
feature, which can be seen to span the entire nuclear region. The gas at 
the lower edge of the streamer, in the 20 km sec$^{-1}$ cloud, also shows 
an increase in line width.  This is the region where we discussed in section 
3.4 a possible disruption of the gas by a nearby supernova remnant. \newline

\subsection{Derived Parameters}

\ \ \ \ \ We derive several physical parameters for the gas along
the streamer, where the signal to noise is high and there
is good agreement between the NH$_3$(1,1) and NH$_3$(2,2) maps.  We also 
compare our derived parameters with those found by Okumura et al. 
(1989) for the northern part of the 20 km sec$^{-1}$ cloud. 
Optical depths can be directly measured from the spectra, using the 
relative intensity of the main and hyperfine lines.
The optical depth of the main line, $\tau$$_m$(J,K), is derived with
the following equation (see \cite{hoe83}), 
\begin{equation}
\frac {\Delta T_a^{*}(J,K,m)}{\Delta T_a^{*}(J,K,s)} = \frac {1-exp[-\tau_m
(J,K)]}{1-exp[-\tau_s(J,K)]},
\end{equation}
where $\Delta$$T_a$$^{*}$ is the observed brightness temperature, {\it m} 
and {\it s} refer to the main and satellite components, and 
$\tau$$_s$(J,K)=$a\tau$$_m$(J,K) is the optical depth of the satellite,
where $a$ is the known ratio of the intensity of the satellite compared with 
the main component. This equation assumes equal beam-filling factors 
and excitation temperatures for the different hyperfine components.  
\newline

\ \ \ \ \ We derive optical depths for the NH$_3$(1,1) transition in each of 
the large cloud clumps along the streamer where hyperfine lines are clearly 
seen in NH$_3$(1,1) spectra.  We will refer to the
 northernmost clump, located at the southeast edge of the CND where the 
streamer ends to the north, as the northern cloud. The 
roughly spherical clump (as seen in NH$_3$(1,1)) directly below the 
northern cloud we will refer to as the 
central cloud, while the larger elongated southern half of the 
steamer will be called the southern cloud.  The southern 
cloud has two peaks of emission, and this cloud corresponds to the
northern part of the 20 km sec$^{-1}$ cloud.  We derive optical depths 
for 44 locations in the southern cloud, 28 locations in the central cloud
and 10 locations in the northern cloud, as seen in Figure \ref{fig.Trot}.
The individual optical depths are reported in Table \ref{rottab}, while
the mean values are shown in Table \ref{tab1}, where we derive the mean for the northern
cloud using only the 5 optically thick estimates of the optical depth.
The error bars for the optical
depth calculations vary across the cloud, as the signal to noise changes.
Near the peak emission of each cloud, where the signal to noise is highest,
the error bars are roughly $\pm$0.2, whereas at the edge of the cloud
the error bars are $\pm$0.9.  Our optical depths for the lower edge of the 
southern cloud agree well with the those derived for the 20 km 
sec$^{-1}$ cloud by Okumura et al. (1989). \newline

\ \ \ \ \ Rotation temperatures for the
gas can be derived if two or more NH$_3$ transitions are observed.  The ratio of
the observed brightness temperatures from the NH$_3$(1,1) 
and NH$_3$(2,2) transitions are related to the rotation temperature as
follows (Equation 4 in \cite{hoe83}),
\begin{eqnarray}
T_R(2,2;1,1) = -41.5 \div \ln\left[ 
\frac{-0.282}{\tau_m(1,1)} \times \right. \ \ \ \ \ \ \ \nonumber \\
\left.\ln\left( 1 - \frac{\Delta T_a^{*}(2,2,m)}{\Delta T_a^{*}(1,1,m)} 
\times (1-e^{-\tau_m(1,1)}) \right)\right]. \end{eqnarray}  
This equation assumes equal beam-filling factors 
and excitation temperatures for the NH$_3$(1,1) and NH$_3$(2,2) transitions.
We calculate rotation temperatures for the locations 
where we have derived optical depths and report the values in Table \ref{rottab}.  
For both the southern and central clouds we find a mean rotation temperature of 22 K,
while for the northern cloud the mean is 32 K.  Error bars on these mean values are
difficult to determine, as the equation is non-linear and the rotation temperature
is highly sensitive to the ratio of the NH$_3$(2,2)/NH$_3$(1,1) brightness temperatures.
From the spread of the values of $T_R$(2,2:1,1) reported in Table \ref{rottab}, we estimate
rough error bars of -5K and +30K for $T_R$(2,2:1,1). 
Our values for $T_R$ for the southern cloud are consistent with those found by
Okumura et al. (1989). \newline

\ \ \ \ \ From the rotation temperature we can estimate a gas kinetic
temperature, $T_K$, using the $T_K$ - $T_R$ relation of Danby et al. (1988).
This $T_K$ - $T_R$ relation is calculated for a `standard' cloud and 
is not very sensitive to cloud density or 
ammonia abundance, however, relations like these may be highly model 
dependent. For the southern and central clouds, where $T_R$ is roughly
15 K to 25 K, the corresponding values for $T_K$ are 17 K to 35 K. For the northern
cloud, where $T_R$ varies from 15 K to 60 K,  $T_K$ ranges from 17 K up to
300 K, where $T_K$ becomes asymptotically large as $T_R$ approaches 60 K.   
Our estimates of the gas temperature for the southern cloud agree well with 
values reported by others for the 20 km sec$^{-1}$ cloud, which range
from 20 K to 120 K (\cite{oku89} and references therein).  The kinetic gas 
temperature in the 20 km sec$^{-1}$ cloud seems to be greater than the dust 
temperature (\cite{mez86}; \cite{zyl90}), so that there may be a difference
in how the dust and gas are heated.  There are some locations where we can not 
derive rotation temperatures. In these cases, the NH$_3$(1,1) optical depth 
is large and the $\Delta$ $T_a^{*}(2,2,m)$ 
to $\Delta$ $T_a^{*}(1,1,m)$ ratio is greater than one. In the optically
thick case, $T_R$ becomes asymptotically sensitive to this brightness temperature ratio,
and we believe that noise is our limiting factor for these few cases where
we are unable to derive $T_R$.  \newline

\ \ \ \ \ The excitation temperature of the gas is derived using 
the following standard radiative transfer equation,
\begin{equation}
\Delta T_a^{*} = \Phi[J_{\nu}(T_{ex}) - J_{\nu}(T_{bg})](1-e^{-\tau}),
\end{equation}
where $T_{ex}$ is the excitation temperature, $T_{bg}$ is the background 
temperature, and $J_{\nu}$ is the Planck function. The brightness and 
excitation
temperatures for the peak emission in each cloud is shown in Table \ref{tab1},
where the error bars on $\Delta$$T_a$$^{*}$ are roughly $\pm$0.2 K, and the 
error bars on $T_{ex}$ are $\pm$0.4-1.0 K. In the equation used to derive
$T_{ex}$, $\Phi$ is the beam-filling 
factor, which is $\sim$1 for an extended source but may be $<$1 
if there exists structure in the clouds on a scale smaller than
the synthesized beam.  There is some indication that full-resolution 
maps of our data (beam size 3.8$''$$\times$1.7$''$) show small structures at 
the peaks of the clouds; therefore the beam-filling factor may be $<$1 and 
our values for $T_{ex}$ may be underestimated.\newline

\ \ \ \ \ The column density, $N$(J,K), can be derived from the observed 
optical depth and excitation temperature. Townes \& Schawlow (1955) 
relate $\tau$, $T_{ex}$
and the column density in the upper level of the transition, $N_1$, as
\begin{equation}
\tau(\nu) = \frac{c^2 h A_{1-0} f(\nu) N_1}{8 \pi \nu k J_{\nu}(T_{ex})},
\end{equation}
where $A_{1-0}$ is the Einstein coefficient for spontaneous emission (1.67
 10$^{-7}$ s$^{-1}$ for (J,K)=(1,1) and 2.23 10$^{-7}$ for 
(2,2)) and $f(\nu$) = (4$\ln$2/$\pi$)$^{1/2}$($\Delta$$\nu$)$^{-1}$ is the 
line profile function for a Gaussian.  We then assume that 
$N$(J,K)=2$N_1$(J,K), which should be valid because the inversion doublet is 
only separated by $\sim$ 1K 
so that any excitation should equilibrate between the two levels.   
Using the usual partition function, with $T_R$=20K-30K, we 
can estimate the amount of NH$_3$ gas in each transition and determine
$N$(NH$_3$) from $N$(J,K).  If we then assume that $N$(H$_2$)=10$^8 N$(NH$_3$) 
(\cite{seg86}), we can roughly estimate the molecular hydrogen
column density and use the spatial scale of the emission to find the mass
of the streamer.  To estimate these parameters we use the mean optical
depth and peak excitation temperature of each cloud.  The results are shown
in Table \ref{tab1}, where the values for the mass are only rough estimates.  
Our derived parameters for the southern cloud agree well with those
reported by Zylka et al. (1990), who find a mass for the 20 km 
sec$^{-1}$ of $\sim$3 10$^5$ M$_{\solar}$.  The good agreement in masses implies that
our assumed abundances are normal.  They report a column density of 
$\sim$3-8 10$^{23}$ cm$^{-2}$, which is slightly lower than our value.  
 \newline

\ \ \ \ \ It is encouraging that our derived parameters for the southern
cloud agree well with several other reports of the 20 km sec$^{-1}$ cloud.
Studies of NH$_3$, CO, and dust emission in the region all yield similiar
numbers, indicating that there are not unusual chemical abundance effects
present and that the derived numbers are robust.  NH$_3$ emission in 
particular is able to clearly detect discrete sources and dense cloud
clumps in this region, whereas often CO, HCN, and dust maps include fluffy,
extended emission which confuses source structure, especially on fine
scales. \newline

\section{Discussion}

\ \ \ \ \ Our results agree well with the general morphology found by many 
other studies of the Galactic Center. The morphology of the 
streamer as mapped in NH$_3$(3,3) (\cite{hoe91}) is similiar to 
our NH$_3$(1,1) and NH$_3$(2,2) maps.  We find little or no emission 
in NH$_3$(1,1) and NH$_3$(2,2) at the location of Sgr A$^*$ and in the central
parsec (Figure \ref{fig.bw.mom0.spiral}).  Ho et al. (1991) also find a 
1.5-2 pc cavity surrounding Sgr A$^*$ 
with no NH$_3$(3,3) emission in the nuclear region, and Jackson et al. (1993)
confirm that molecular gas is not found in this cavity. 
The key results in this paper concern the approach of 
molecular material toward this central cavity.   \newline

\subsection{The Case for Infalling Gas}

\ \ \ \ \ Ho et al. (1991) discuss the possibility of gas feeding the 
CND from the 20 km s$^{-1}$ cloud via the southern streamer.  Other studies
have since imaged the same streamer, but none
have found any kinematical or direct evidence to support the suggestion of 
infall or accretion  (Marr et al. 1995; \cite{den93}).
Whether this gas streamer is seen only in projection against the Galactic 
Center or whether it is actually approaching the central mass concentration
can only be determined if we detect direct effects of the deepening 
gravitational potential well on the purported infalling feature.
Our new data support this accretion theory with morphological, thermal, and
kinematic evidences: \newline

\ \ \ \ \ (1) Our velocity-integrated maps show a long, narrow 
streamer connecting the 
northern part of the 20 km sec$^{-1}$ GMC with the CND region (Figure 
\ref{fig.color.mom0}). It is of 
special significance that the streamer $stops$ at the CND. This can
be seen not only in the velocity-integrated map, but also in the 
position-velocity diagrams for cut `a' (Figure \ref{fig.psvl}), 
where the gas moving into the region at 30-35 km sec$^{-1}$ clearly 
stops to the north, while the gas above it has much larger line widths. 
The comparison of the different
NH$_3$ transitions suggests that the termination of the streamer is $not$
a temperature effect. The morphology
of the streamer in NH$_3$(1,1) and NH$_3$(2,2) agrees well with NH$_3$(3,3),
HCN and sub-millimeter continuum studies.   \newline

\ \ \ \ \ (2) We find that the northern tip of the streamer, located
nearest to Sgr A$^*$ in projection, is stronger in NH$_3$(2,2) emission than 
in NH$_3$(1,1), indicating that the gas is heated (Figure 
\ref{fig.22.11mom0}).  Other locations of heating appear around 
the CND. This would seem to indicate the gas is located at the distance
of the Galactic Center and is being heated by nuclear processes.  \newline

\ \ \ \ \ (3) A velocity gradient is detected along the streamer.
In the northern part of the streamer we observe a gradient of 
$\sim$5 km sec$^{-1}$ arcmin$^{-1}$ (Figure \ref{fig.psvl}). Along the 
southern half of the 
streamer (located in declination from -29$^\circ$01$'$ to -29$^\circ$03$'$) 
the gradient is
 $\sim$5-8 km sec$^{-1}$ arcmin$^{-1}$ seen across $\sim$5 pc. 
This is in good agreement with the southwest to northeast gradient of
5 km sec$^{-1}$ arcmin$^{-1}$ seen in the 20 km sec$^{-1}$ cloud
by Okumura et al. (1989) and the 11 km sec$^{-1}$ arcmin$^{-1}$ gradient
seen by Zylka et al. 1990.  The expected gradient at a radius of 5 pc 
for a central mass
of 10$^7$ M$\solar$ (including the central compact source and stellar
population, see \cite{mor96}; \cite{hal96}) is $\sim$9 km sec$^{-1}$ pc$^{-1}$, 
roughly 3 or 4 times more than 
we observe. Thus, our detected gradient suggests that if this motion is 
infall, the bulk of the motion is across the line
of sight, with the geometry such that the streamer is more to the side of
Sgr A$^*$ than along our particular line of sight.  The difference in 
the size of the gradient seen in the northern part of the streamer as
opposed to the southern part may indicate a tilting in the streamer, so that 
projection effects may result in the
lower gradient seen to the north.   There may 
also be forces other than gravity which need to be taken into account, 
such as increased turbulence from supernovae and winds as well as magnetic
effects. \newline

\ \ \ \ \ (4) Another handle on the kinematics is the velocity dispersion of 
the observed motions. 
The streamer ends to the north at the location of the CND, where we see in the 
position-velocity diagrams that the line width increases dramatically (Figures 
\ref{fig.11.spec}, \ref{fig.22.spec}, \ref{fig.psvl}).  This increase
in line width has not been seen before.  Other studies of the southern 
streamer have coarser spectral resolutions than our 5 km sec$^{-1}$.  
It is expected 
that if the gas is moving from the 20 km sec$^{-1}$ cloud to the nuclear 
region, it would interact with the CND and become disrupted, resulting in 
an increase in the line width of the gas, regardless of whether
the motion is predominantly along or across the line of sight.  
The disk is known to have inherently large line widths around $\geq$ 40 km sec$^{-1}$.  
Our detection and measurement of a component with increased line widths 
concretely place the northern tip of the streamer at the location of the CND 
and are consistent with the gas in the southern streamer accreting onto 
the CND.  This kinematic evidence does not reflect chemistry or abundance 
effects and is the strongest argument for the streamer transporting gas 
to the nuclear region along the southern streamer.  The morphology and
velocity gradient we see do not argue alone that the streamer is
falling in towards the nuclear region; it is only when combined with
the heating seen and the increase in line width at the northern tip 
of the streamer that the argument becomes strong.
\newline

\subsection{Other Nearby Streamers?}

\ \ \ \ \ Hints in our data show that there may be more than
one streamer flowing toward the nuclear region.  The 20 km 
sec$^{-1}$ cloud has a tuning-fork morphology which may 
be indicating the 
presence of two streamers.  The southern edge of streamer which dips into
the 20 km sec$^{-1}$ cloud has two separate elongated lobes of emission.
These two `fingers' pointing into the GMC can be seen clearly in the
 velocity dispersion maps (Figure \ref{fig.mom2}), where the dark 
regions at the southern 
part of the map seem to separate into two distinct regions of gas.
There are also hints in the velocity-integrated maps 
(Figure \ref{fig.color.mom0} and the NH$_3$(2,2) contours in Figure \ref{fig.22.11mom0}) 
that there may be a second streamer connecting
the northern section of this double-lobed part of the 20 km sec$^{-1}$ 
cloud to the nuclear region.  Along the western edges of the maps
(e.g. RA=17$^h$42$^m$26$^s$)
there are hints of a narrow line of gas originating from the northwest of
the 20 km sec$^{-1}$ cloud and moving up towards the southwestern
quadrant of the CND.  Both this second western streamer
as well as the main southern streamer have been imaged in dust emission
 at 1.3mm by Zylka 
et al. (1997), where their maps show remarkable agreement with our NH$_3$ emission.
 \newline

\ \ \ \ \ Another feature at the one contour level in the velocity
integrated map which seems to be a real structure is the small spot of
emission to the east of the 20 km sec$^{-1}$ seen in both
the NH$_3$(1,1) and NH$_2$(2,2) maps (RA=17$^h$42$^m$32$^s$).  
Spectra from this location
indicate a central velocity of $\sim$15-25 km sec$^{-1}$.  
It is also worth noting that this feature is seen clearly
in the NH$_3$(3,3) data of Ho et al. (1991), where it is
part of a ridge of gas connecting the 20 km sec$^{-1}$ with the
50 km sec$^{-1}$ cloud. The higher intensity of
this feature in the NH$_3$(3,3) line suggests that it must be
fairly warm.   \newline

\section{Conclusions}

\ \ \ \ \ Investigating connections between molecular material in the nucleus
of the Galaxy and the nearby 20 km sec$^{-1}$ GMC, NH$_3$(1,1) and NH$_3$(2,2) 
were observed with the VLA using
three overlapping 2$'$ fields at the Galactic Center, surrounding Sgr A$^*$
and the region directly south of it.
We map a streamer which appears to be feeding the nuclear region with 
molecular gas from the 20 km sec$^{-1}$ cloud.  
The long, narrow streamer is about 2 pc east to west and 10 
pc north to south and originates
from the northeastern edge of the 20 km sec$^{-1}$ cloud, south of Sgr A West.  
The streamer ends to the north
at the location of the CND, in the area where the least amount of
HCN disk flux is seen, just southeast of Sgr A$^*$.  
We detect a clear velocity gradient along its length of $\sim$5-8 km sec$^{-1}$
arcmin$^{-1}$.  The line widths of the gas increase substantially
in the northern portion of the streamer, with FWHMs of $\geq$50 km
 sec$^{-1}$, possibly signalling accretion
onto the CND.  The gas at the location of the CND also appears to be heated, 
as indicated by the ratio of the NH$_3$(2,2)/NH$_3$(1,1) velocity-integrated
emission.  Additional faint streamers also appear to be link the 20 km sec$^{-1}$
cloud to the CND, suggesting possible other paths of accretion.
Our morphological, kinematical, and thermal data strongly support the thesis 
that gas is falling in towards the circumnuclear region from the
20 km sec$^{-1}$ cloud along the southern streamer and accreting onto the CND.
 \newline

\ \ \ \ \ We would like to thank Mel Wright for providing the HCN maps
of the CND. The image of the ionized gas spiral in Sgr A West was made by
 D. A. Roberts and W. M. Goss and was downloaded from the Astronomy Digital 
Image Library maintained by the NCSA.  
This work was supported in part by the National Science Foundation 
under the auspices of the REU program at the Harvard-Smithsonian Center for 
Astrophysics. \newline

\newpage

\figcaption[fig_mom0.a+as+1/color.mom0.cps]{Velocity-integrated maps of 
NH$_3$(1,1) and NH$_3$(2,2) emission at the Galactic Center are shown 
in yellow contours overlaid on a coded intensity color HCN image of the 
circumnuclear 
disk (CND). A gas streamer can be seen to the south of 
the CND, while a semi-complete ring of NH$_3$ emission closely follows 
the CND itself.  Three 2$'$ 
fields have been mosaiced together, with the overlapping primary beams 
shown as a dotted contour. These images have significant primary beam 
attenuation at the edges, and the noise level across the central region
of the image is .05 Jy beam$^{-1}$ km sec$^{-1}$.
The solid contours are at integer levels of .4 Jy beam$^{-1}$ km sec$^{-1}$,
and the 14$''$$\times$9$''$ synthesized beam is in the lower left corner.
\label{fig.color.mom0}}

\figcaption[fig_mom0.a+as+1/spiral.bw.mom0.ps]{Northern two fields of the 
velocity-integrated 
NH$_3$(1,1) and NH$_3$(2,2) emission in contours overlaid on a greyscale 
image of the mini-spiral of ionized gas surrounding Sgr A$^*$.  
The contour levels are the same as in Figure \ref{fig.color.mom0}. 
\label {fig.bw.mom0.spiral}}

\figcaption[fig_22.11mom0.a+as+1/22.11.mom0.both.ps]{NH$_3$(2,2) 
primary-beam corrected emission
is shown in contours overlaid on the NH$_3$(2,2)/NH$_3$(1,1) 
line intensity ratio in greyscale.  The NH$_3$(2,2)/NH$_3$(1,1) line ratio
indicates heating, where the darker areas correspond to hotter gas.  
The hottest area is at the northern tip of the streamer, where it 
reaches the CND. The greyscales are 0 to 50 Jy beam$^{-1}$ km sec$^{-1}$ 
for the NH$_3$(2,2) emission and a ratio of 0.5 to 6 for the divided
NH$_3$(2,2)/NH$_3$(1,1) emission. \label{fig.22.11mom0}}

\figcaption[fig_heat.spiral.a+as+1/heat.spiral+disk.ps]{The
NH$_3$(2,2)/NH$_3$(1,1) line intensity ratio (the same as in Figure 
\ref{fig.22.11mom0}) is shown in greyscale with contours of the 
ionized gas mini-spiral on the left and the CND as seen in HCN on the right. 
\label{fig.22.11heat}}

\figcaption[fig_11spec.a+as/11.spec.fig.ps]{Selected NH$_3$(1,1) 
spectra with Gaussian fits; the positions of the spectra are 
shown in the primary-beam corrected map. The line widths of the spectra
increase as the gas is located near the CND.  \label{fig.11.spec}}

\figcaption[fig_22spec.a+as/22.spec.fig.ps]{Selected NH$_3$(2,2) 
spectra with Gaussian fits from a primary-beam corrected map. \label{fig.22.spec}}

\figcaption[fig_basespec.a+as/base.spec.ps]{Test of baseline subtraction 
using random NH$_3$(1,1) and NH$_3$(2,2) spectra from low-emission regions;
there is no residual emission above zero.
\label{fig.base.spec}}

\figcaption[fig_psvl.a+as+1/22.11.psvl.ps]{Position-velocity diagrams 
for NH$_3$(1,1), NH$_3$(2,2) and 
NH$_3$(2,2)/NH$_3$(1,1) emission; the second row for the cut 
`a' has been smoothed in the R.A. direction, while the other diagrams have
no smoothing applied. North is up in the `a' and `b' diagrams,
while east is up for the `c' diagrams. 
The NH$_3$(2,2)/NH$_3$(1,1) greyscale ranges from a ratio
of 0.5 (light) to 3 (dark) with contours at 0.5 intervals, and the 
contour levels for the NH$_3$(1,1) and NH$_3$(2,2) diagrams 
are at .2 Jy beam$^{-1}$.  \label{fig.psvl}}

\figcaption[fig_mom0.a+as+1/bw.mom2.ps]{Velocity dispersion maps which 
trace line width gradients in the gas, where the darker areas indicate 
broader line widths.  The line widths increase in the nuclear area in
both the NH$_3$(1,1) and NH$_3$(2,2) gas, as well as at the southern 
tip of the long, narrow southern streamer. The greyscale is 0 km 
sec$^{-1}$ (white) to 40 km sec$^{-1}$ (dark). 
 \label{fig.mom2}}

\figcaption[fig_mom0.a+as+1/bw.Trot.ps]{The locations used to derive
optical depths and rotation temperatures listed in Table 1 are labelled on a 
grey-scale velocity-integrated map of the NH$_3$(1,1) emission.  \label{fig.Trot}}

\pagebreak

\small
\begin{deluxetable}{ccccc}
\tablewidth{350pt}
\tablecolumns{5}
\tablecaption{Derived Physical Parameters \label{rottab}}
\tablehead{
\colhead{Location}\tablenotemark{a} & \colhead{$\tau$(1,1,m)} &
\colhead{$\Delta$ $T_a^{*}(1,1,m)$} & 
\colhead{$\Delta$ $T_a^{*}(2,2,m)$ } &
\colhead{$T_R$(2,2:1,1)} \\
\colhead{} & \colhead{$(K)$} & \colhead{$(K)$}
& \colhead{$(K)$} }
\startdata
A & $\ll$1 & 1.5 & 2.7 & 58 \\
B & 3.4  & 0.9 & 1.4 & \ \ $ ^b$    \\
C & $\ll$1 & 1.9 & 2.7 & 46 \\
D & $\ll$1 & 2.5 & 2.7 & 36\\
E & $\ll$1 & 1.9 & 1.6 & 29\\
F & 2.3 & 1.7 & 1.7 & 34 \\
G & 1.3 & 2.5 & 2.1 & 26 \\
H & 1.4 & 2.7 & 2.0 & 23 \\
I & 3.7 & 1.9 & 1.5 & 19 \\
J & 5.6 & 0.6 & 0.4 & 15 \\
K & 1.7 & 0.9 & 0.7 & 21  \\
L & 1.9 & 1.5 & 1.1 & 22  \\
M & 2.2 & 1.4 & 1.4 & 36  \\
N & 1.9 & 1.4 & 1.8 & \ \ $ ^b$     \\
O & 2.0 & 0.9 & 0.9 & 32  \\
P & 2.1 & 1.6 & 1.5 & 26  \\
Q & 2.2 & 2.4 & 1.9 & 23  \\
R & 2.5 & 2.4 & 2.0 & 23  \\
S & 1.3 & 2.2 & 1.7 & 25  \\
T & 2.0 & 2.0 & 2.0 & 32  \\
U & 2.3 & 3.2 & 2.7 & 24  \\ 
V & 1.6 & 3.4 & 2.5 & 22 \\ 
W & 0.9 & 3.0 & 1.9 & 22  \\ 
X & 1.7 & 3.5 & 2.6 & 23  \\ 
Y & 2.2 & 3.8 & 2.6 & 20  \\ 
Z & 3.4 & 3.5 & 2.2 & 16  \\ 
a & 2.9 & 2.9 & 2.1 & 19  \\ 
b & 4.0 & 3.4 & 2.6 & 18  \\ 
c & 5.8 & 3.4 & 2.6 & 15  \\ 
d & 5.0 & 3.2 & 2.3 & 15  \\ 
e & 3.3 & 2.4 & 1.5 & 17  \\ 
f & 2.0 & 1.1 & 0.6 & 17  \\ 
g & 3.0 & 1.7 & 1.0 & 17  \\ 
h & 1.8 & 2.4 & 1.7 & 21  \\ 
i & 1.9 & 3.0 & 2.7 & 27  \\ 
j & 2.9 & 3.2 & 3.1 & 29  \\ 
k & 0.3 & 2.9 & 2.6 & 31  \\ 
l & 1.3 & 2.2 & 1.5 & 22  \\ 
m & 3.1 & 1.2 & 0.8 & 18  \\ 
n & 6.3 & 0.8 & 0.8 & 21 \\ 
o & 4.0 & 1.1 & 0.8 & 16 \\ 
p & 4.1 & 1.1 & 0.8 & 16 \\ 
q & 4.2 & 1.3 & 1.1 & 19 \\ 
r & 4.1 & 1.5 & 1.0 & 16 \\ 
s & 6.0 & 1.4 & 0.9 & 14 \\ 
t & 6.2 & 1.0 & 0.9 & 18 \\ \tablebreak
u & 9.8 & 0.8 & 0.8 & 19 \\ 
v & 7.1 & 1.4 & 0.9 & 13 \\ 
w & 3.7 & 1.9 & 1.7 & 24 \\ 
x & 7.2 & 1.7 & 1.9 & \ \ $ ^b$  \\ 
y & 6.2 & 1.9 & 2.3 & \ \ $ ^b$  \\ 
z & 4.3 & 1.1 & 1.3 & \ \ $ ^b$  \\ 
$A$ & 6.1 & 1.2 & 1.2 & 20 \\ 
$B$ & 4.4 & 2.2 & 2.3 & \ \ $ ^b$   \\ 
$C$ & 4.2 & 2.2 & 2.4 & \ \ $ ^b$   \\ 
$D$ & 4.8 & 2.3 & 2.5 & \ \ $ ^b$   \\ 
$E$ & 3.6 & 2.6 & 2.6 & 29  \\ 
$F$ & 6.1 & 1.5 & 2.1 & \ \ $ ^b$    \\ 
$G$ & 5.1 & 2.4 & 2.0 & 17 \\ 
$H$ & 4.0 & 2.3 & 2.3 & \ \ $ ^b$  \\ 
$I$ & 3.9 & 3.8 & 2.9 & 18 \\ 
$J$ & 1.8 & 4.4 & 5.6 & \ \ $ ^b$   \\ 
$K$ & 1.9 & 5.0 & 6.3 & \ \ $ ^b$   \\ 
$L$ & 3.4 & 3.7 & 2.3 & 16 \\ 
$M$ & 2.3 & 3.6 & 3.0 & 23 \\ 
$N$ & 3.3 & 4.4 & 3.1 & 18 \\ 
$O$ & 2.5 & 4.8 & 4.6 & 29 \\ 
$P$ & 2.3 & 5.3 & 5.0 & 29 \\ 
$Q$ & 1.7 & 5.0 & 4.6 & 28 \\ 
$R$ & 1.3 & 2.4 & 2.7 & 39 \\ 
$S$ & 1.2 & 2.1 & 2.3 & 40 \\ 
$T$ & 2.2 & 3.1 & 3.4 & 60 \\ 
$U$ & 2.9 & 2.1 & 2.1 & 34 \\ 
$V$ & 2.3 & 3.0 & 2.5 & 23 \\ 
$W$ & 2.1 & 3.2 & 3.8 & \ \ $ ^b$   \\ 
$X$ & 0.9 & 1.7 & 2.1 & 48 \\ 
$Y$ & 0.8 & 1.6 & 1.7 & 35 \\ 
$Z$ & 0.9 & 1.4 & 1.2 & 26 \\ 
$a$ & 0.8 & 1.3 & 1.2 & 28 \\ 
$b$ & 1.5 & 1.0 & 1.2 & 61 \\ 
$c$ & 0.8 & 1.5 & 1.6 & 34 \\ 
$d$ & 1.2 & 1.4 & 1.8 & 90 \\ 
$e$ & 1.2 & 1.5 & 1.8 & 53 \\ 
\enddata
\tablenotetext{a}{The positions for each location listed here are shown
in Figure 10}
\tablenotetext{b}{Rotation temperature can not be derived (see section 3.6 for 
discussion)}  
\end{deluxetable}

\begin{deluxetable}{lcccccc}
\tablecolumns{7}
\tablecaption{NH$_3$(1,1) Column Density and Mass Estimates \label{tab1}} 
\tablehead{
\colhead{Cloud} & \colhead{Mean $\tau$$_m$} &
\colhead{Peak $\Delta$$T_a$$^{*}$} & \colhead{Peak $T_{ex}$} &
\colhead{$N(J,K)$} & \colhead{$N$(H$_2$)} & \colhead{Mass}  \\
\colhead{} & \colhead{} & \colhead{$(K)$} & \colhead{$(K)$} 
& \colhead{($10^{15}$ $cm^{-2}$)} & \colhead{($10^{24}$ $cm^{-2}$)} & 
\colhead{($10^4$ $M_{\solar}$)} }
\startdata
Northern & 2.7 & 3.2 & 6.1 & 3.4 & 0.85 & 5.5 \\   
Central & 2.4 & 3.8 & 6.9 & 3.4 & 0.81 & 6.2 \\   
Southern & 3.5 & 5.3 & 8.2 & 6.0 & 1.4 & 23    \\   
\enddata
\end{deluxetable}

\end{document}